\documentclass[12pt]{article}

\newcommand{\cem}{{\hspace{1cm}}}
\newcommand{\wz}{{\mbox{{\tiny {WZ}}}}}

\def\ftoday{{\sl  \number\day \space\ifcase\month
\or Janvier\or F\'evrier\or Mars\or avril\or Mai
\or Juin\or Juillet\or Ao\^ut\or Septembre\or Octobre
\or Novembre \or D\'ecembre\fi
\space  \number\year}}

\makeatletter
\@addtoreset{equation}{section}
\makeatother

\newcommand{\ta}{{\mbox{\tiny{A}}}}
\newcommand{\tb}{{\mbox{\tiny{B}}}}
\newcommand{\tc}{{\mbox{\tiny{C}}}}
\newcommand{\td}{{\mbox{\tiny{D}}}}
\newcommand{\te}{{\mbox{\tiny{E}}}}
\newcommand{\tf}{{\mbox{\tiny{F}}}}

\renewcommand{\a}{\alpha}
\renewcommand{\b}{\beta}
\newcommand{\g}{\gamma}           
\renewcommand{\d}{\delta}         
\newcommand{\e}{\epsilon}

\newcommand{\la}{\lambda}

\newcommand{\p}{\psi}             

\newcommand{\s}{\sigma}           
         
\newcommand{\f}{{\phi}}           \newcommand{\F}{{\Phi}}
\newcommand{\vf}{{\varphi}}
\newcommand{\x}{\xi}              

\newcommand{\z}{\zeta}

\newcommand{\eps}{{\epsilon}}

\newcommand{\ca}{{\cal A}}
\newcommand{\cb}{{\cal B}}
\newcommand{\cc}{{\cal C}}
\newcommand{\cd}{{\cal D}}

\newcommand{\cf}{{\cal F}}

\newcommand{\cm}{{\cal M}}
\newcommand{\cn}{{\cal N}}

\newcommand{\br}[1]{\left({#1}\right)}
\newcommand{\brg}[1]{\left({#1}\right)_{T}}
\newcommand{\brh}[1]{\left({#1}\right)_{H}}

\newcommand{\be}{\begin{equation}}
\newcommand{\ee}{\end{equation}}
\newcommand{\eqn}[1]{\label{#1}\end{equation}}

\newcommand{\bea}{\begin{eqnarray}}
\newcommand{\eea}{\end{eqnarray}}
\newcommand{\eqan}[1]{\label{#1}\end{eqnarray}}

\newcommand{\ba}{\begin{array}}
\newcommand{\ea}{\end{array}}

\newcommand{\loco}{{\mathop{ \, \rule[-.06in]{.2mm}{3.8mm}\,}}}
\newcommand{\doubar}{{{\loco}\!{\loco}}}

\newcommand{\nn}{\nonumber}
\newcommand{\equ}[1]{(\ref{#1})}

\newcommand{\bp}{\bar{\psi}}

\newcommand{\bs}{{\bar{\sigma}}}

\newcommand{\bl}{\bar{\lambda}}

\newcommand{\au}{{\bf u}}
\newcommand{\av}{{\bf v}}
\newcommand{\aw}{{\bf w}}

\newcommand{\da}{{\dot{\alpha}}}
\newcommand{\db}{{\dot{\beta}}}
\newcommand{\dg}{{\dot{\gamma}}}
\newcommand{\dd}{{\dot{\delta}}}
\newcommand{\df}{{\dot{\varphi}}}

\begin{document}

\font\fifteen=cmbx10 at 15pt \font\twelve=cmbx10 at 12pt

\begin{titlepage}

\begin{center}

\renewcommand{\thefootnote}{\fnsymbol{footnote}}

{\twelve Centre de Physique Th\'eorique\footnote{Unit\'e Propre de
Recherche 7061
}, CNRS Luminy, Case 907}

{\twelve F-13288 Marseille -- Cedex 9}

\vspace{3 cm}

{\fifteen  N=4 Supergravity with Antisymmetric Tensor\\[2mm] in Central Charge
Superspace}

\vspace{1.4 cm}

{\bf Richard GRIMM$\,^1$, Carl HERRMANN$\,^2$ and Annam\'aria KISS$\,^1$ }
\vspace{1cm}

$^1${\em Centre de Physique Th\'eorique, CNRS Luminy Case 907,\\ F--13288
Marseille Cedex 9, France} \bigskip

$^2${\em Fachgruppe Theoretische Physik - FB Physik, Martin Luther Universit\"at
Halle--Wittenberg,  D--06099  Halle, Germany}

\vspace{1.5cm}

{\bf Abstract} \end{center}

\vspace{0.3cm}

A concise geometrical formulation of $N=4$ supergravity containing an
antisymmetric tensor gauge field is given in central charge superspace:
graviphotons are identified in the super--vielbein on the same footing as
the vierbein and the Rarita--Schwinger fields.
As a consequence of superspace soldering,  Chern--Simons terms in the
fieldstrength
of the antisymmetric tensor arise as an intrinsic property of superspace with
central charge coordinates.

\vspace{1cm}

\vfill


\bigskip

\bigskip

\noindent September 2000

\noindent CPT-2000/P.4027

\bigskip

\noindent anonymous ftp : ftp.cpt.univ-mrs.fr

\noindent web : www.cpt.univ-mrs.fr

\end{titlepage}

\section{Introduction}

$N=4$ is the smallest extended supergravity theory which contains on--shell
helicity $0$ states. There are two versions of $N=4$ supergravity which realise
both of the two helicity $0$ states as a complex scalar: one with an
off--shell $SO(4)$ and an on--shell $SU(4)$ global symmetry
and another one with a global off-shell $SU(4)$ symmetry
\cite{Das77}, \cite{CS77}, \cite{CSF77}, \cite{CSF78}, \cite{FS78},
\cite{Gri78}.
A third version, containing a real scalar and an antisymmetric
tensor gauge field as helicity $0$ states in its spectrum \cite{GSO77}, has been
formulated in terms of component fields by Nicolai and Townsend \cite{NT81}.
The latter, which we will call the N--T multiplet in the sequel,
is related to the previous
theory by a scalar-tensor duality transformation.

As to geometric formulations in superspace \cite{Sie81}, \cite{How82},
\cite{Gat83}, the former, containing a complex scalar with $SU(1,1)/U(1)$
coset structure, has been given both in canonical superspace \cite{Gat83} and
in superspace extended by bosonic coordinates corresponding to central charges
\cite{How82}.  On the other hand, the superspace formulation of the N--T
multiplet encountered a number of problems, which were identified in
\cite{Gat83}, and overcome in \cite{GD89} in introducing external Chern--Simons
forms for the graviphotons.  The present approach differs from \cite{GD89} in
that graviphoton Chern--Simons forms appear intrinsically as a property of
central charge superspace and a consequence of the superspace soldering
mechanism.  This concise central charge superspace formulation allows to derive
the N--T theory in a natural and straightforward way.

Recall that in central charge geometry, the frame of superspace $E^\ca$ has
components $E^\ca=(E^a,E_\ta^\a,E^\ta_\da, E^\au)$. Here $a$, $\a$, $\da$ denote
the usual vector and Weyl spinor indices while capital indices $\ta$ count the
number of supercharges and boldface indices $\au$ the number of central
charges. This
framework provides a unified geometric identification of graviton,
gravitini {\em
and} graviphotons as
lowest superfield components such that
\be E^a\doubar\ =\ dx^m e_m{}^a~,\quad
E_\ta^\a\doubar\ =\ \frac{1}{2}\,dx^m \p_{m \, \ta}^{ \ \ \a}~, \quad
E^\ta_\da\doubar\ =\ \frac{1}{2}\,dx^m \bar{\p}_{m \, \da}^{\ \ \ta}~, \quad
E^\au\doubar\ =\ dx^m v_m{}^\au~.
\eqn{frame}
The double bar projects at the same time on the vector
coefficient of the differential form and on the lowest superfield component.
On the other hand, the antisymmetric tensor is identified in a superspace
2--form
$B$ such
that
\be B\doubar\ =\ \frac{1}{2}\,dx^m dx^n b_{nm}. \eqn{B}

At this stage we still have two separate geometric structures: the supergravity
sector and the 2--form sector. In order to identify $b_{mn}$ in the same
multiplet as
$e_m{}^a$,
$\ \p_{m \, \ta}^{ \ \ \a}$,
$\ \bar{\p}_{m \, \da}^{\ \ \ta}$,
$v_m{}^\au$
we will use a mechanism called superspace
soldering, already known in other contexts of superspace geometry. This
mechanism
relies crucially on the special form of certain torsion coefficients. In
our case
they are
\be
T{^\tc_\g}{^\tb_\b}{^a_{}}\ =\ 0~,
\cem
T{^\tc_\g}{^\db_\tb}{^a}\ =\
-2i\d{^\tc_\tb}(\s{^a}\eps){_\g}{^\db}~,
\cem
T{^\dg_\tc}{^\db_\tb}{^a}\ =\ 0~,
\eqn{T01} \be T{^\tc_\g}{^\tb_\b}{^\au}\ =\ \eps_{\g\b}T^{[\tc\tb]\au}~,\cem
T^\tc_\g{}^\db_\tb{}^\au\ =\ 0~, \cem T{^\dg_\tc}{^\db_\tb}{^\au}\ =\
\eps^{\dg\db}T{_{[\tc\tb]}}{}^\au~.
\eqn{T02}
As we will show below, the
remaining components of the  multiplet, the real scalar and the helicity 1/2
fields, will be identified in torsion components. Whereas the real scalar will
appear in $T^{[\tc\tb]\au}$ and $T{_{[\tc\tb]}}{}^\au$, the helicity 1/2 fields,
which we shall call {\em gravigini} fields, are identified in the lowest
superfield
components of
\be
T^\tc_\g{}^\tb_\b{}^\ta_\da\ =\ \eps_{\g\b} \, T^{[\tc\tb\ta]}_{\ \ \ \
\da}~,\cem
T_\tc^\dg{}_\tb^\db{}_\ta^\a\ =\ \eps^{\dg\db} \, T_{[\tc\tb\ta]}^{\ \ \ \a}~.
\ee
The scalar, the helicity 1/2 gravigino together with the component fields
defined in
(\ref{frame}) and (\ref{B}) constitute the N--T multiplet.


\section{General geometric superspace structures}

As mentioned above, the basic ingredients in our construction of $N=4$
supergravity are the local frame $E^\ca$ in superspace and the 2--form gauge
potential $B$.

Superspace diffeomorphisms, which now contain general space--time coordinate
transformations, local supersymmetry transformations as well as local
central charge
transformations, are implemented in the usual way. Covariant derivatives with
respect to structure group transformations are defined in terms of the
connection
1--form $\F_\cb{}^\ca$. Torsion, curvature and fieldstrength of the 2--form
gauge
potential are given as
\bea
T^\ca&=&DE^\ca\ =\ dE^\ca+E^\cb \F_\cb{}^\ca~,\\[2mm]
R_\cb{}^\ca&=&d\F_\cb{}^\ca+\F_\cb{}^\cc\F_\cc{}^\ca~,\\[2mm] H&=&dB~.
\eea
Following standard textbook procedures \cite{WB83} we introduce supergravity, or
Wess-Zumino \cite{BGG00} transformations
\bea
\d_\wz E^\ca&=&D\x^\ca+\imath_\x T^\ca~,\label{dE}\\[2mm]
\d_\wz B&=&\imath_\x H~,\\[2mm]
\d_\wz \F_\cb{}^\ca&=&\imath_\x R_\cb{}^\ca~,
\eea
as certain combinations of superspace diffeomorphisms and
structure group transformations. Observe that the complete structure of
commutation relations derives in a constructive way. For later convenience we
display here the explicit form of the transformation of the vielbein and of the
2--form:
\bea
\d_\wz \, E_\cm{}^\ca &=&
\cd_\cm\x^\ca + E_\cm{}^\cb\,\x^\cc T_{\cc\cb}{}^\ca~,
\label{dE_comp}\\[2mm]
\d_\wz \, B_{\cn\cm} &=&
(-)^{n(m+a)}E_{{\cal M}}^{\quad {\cal A}}
E_{{\cal N}}^{\quad {\cal B}}\,\xi ^{{\cal C}}H_{{\cal CBA}}~.
\label{dB_comp}
\eea

Observe that on the level of component fields, {\em i.e.} lowest superfield
values
(\ref{frame}), (\ref{B}), the transformations of parameters
$\xi^\a_\ta\loco=\zeta^\a_\ta$ and
$\xi_\da^\ta\loco=\bar\zeta_\da^\ta\,$ reproduce the Wess--Zumino gauge,
justifying
the notation $\d_\wz$.

In our analysis, we will make systematic use of the torsion and 3--form Bianchi
identities
\be DT^\ca \ =\  E^\cb R_\cb{}^\ca~,\cem dH \ =\  0~. \eqn{BI}
In some more detail, we shall use the notation
\bea &&\brg{_{\cd\cc\cb}{}^\ca}\cem E^\cb
E^\cc E^\cd \br{\cd_\cd T_{\cc\cb}{}^\ca+T_{\cd\cc}{}^\cf
T_{\cf\cb}{}^\ca-R_{\cd\cc\cb}{}^\ca }\ =\ 0~,
\label{BianchiT}\\[2mm]
&&\brh{_{\cd\cc\cb\ca}}\cem E^\ca E^\cb E^\cc E^\cd \br{2\,\cd_\cd
H_{\cc\cb\ca}+3\,T_{\cd\cc}{}^\cf H_{\cf\cb\ca}}\ =\ 0~,\label{BianchiH}
\eea
displaying explicitly the 3--form and 4--form coefficients of the above
relations.



\section{Torsion, 3--form curl and superspace soldering}

Before presenting in detail the superspace geometry which describes the N--T
multiplet, we shall explain qualitatively the mechanism of superspace soldering.
Roughly speaking, this operation allows to identify various components of
one and
the same supergravity multiplet in two distinct geometric structures: graviton
$e_m{}^a$, gravitini
$\ \p_{m \, \ta}^{ \ \ \a}$,
$\ \bar{\p}_{m \, \da}^{\ \ \ta}$,
and graviphotons
$v_m{}^\au$ in the gravity sector, and the antisymmetric tensor $b_{mn}$ in the
2--form sector.

The basic idea is to establish relations between the coefficients of the 3--form
curl $H$ and the torsion $T^\ca$ in suitably choosing constraints in both
sectors. As examples we will describe how $H_{c\,b\,a}$ will appear in the
torsion coefficients, on the one hand, and the relation between
$T_{c\,b}{}^\au$,
the graviphoton fieldstrength and the coefficient $H_{\au\,b\,a}\,$ on the other
hand. This will then lead automatically to the appearance of the graviphoton
Chern--Simons form in the supercovariant component fieldstrength
$H_{c\,b\,a}\loco$.

To see intuitively how the soldering procedure works we will have a look at
certain Bianchi identities. As to the relation of $H_{c\,b\,a}$ to torsions we
consider $\brh{_{\d\tc \, ba}^{\td\dg}}$ in (\ref{BianchiH}). At an early
intermediate stage of the analysis, with some suitably chosen constraints, this
superfield equation takes the form
\bea
&&\cd^\td_\d H_{\tc\,
b\,a}^\dg+ \cd_\tc^\dg H^\td_{\d\, b\,a}+\cd_b H^{\td\,\dg}_{\d\,\tc\,a}-\cd_a
H^{\td\,\dg}_{\d\,\tc\,b}
\nn\\[2mm]&&
+T^{\td\,\dg\,\vf}_{\d\,\tc\,\tf}
H^\tf_{\vf\,b\,a} +T^{\td\,\dg\,\tf}_{\d\,\tc\,\df} H^{\df}_{\tf\,b\,a} +T^{\td\
\, \au}_{\d\,a} H_{\au\,\tc\,b}^{\,\ \dg} +T_{\tc\,a}^{\dg\,\ \au}
H_{\au\,\d\,b}^{\,\ \td} -T^{\td\,\ \au}_{\d\,b}H_{\au\,\tc\,a}^{\,\ \dg}
-T_{\tc\,b}^{\dg\,\ \au} H_{\au\,\d\,a}^{\,\ \td}
\nn \\[2mm]&&
+T^{\td\,\dg\,f}_{\d\,\tc} H_{f\,b\,a} +T^{\td\,\ \vf}_{\d\,a\,\tf}
H^{\tf\,\dg}_{\vf\,\tc\,b} +T_{\tc\,a\,\df}^{\dg\,\ \tf}
H^{\df\,\td}_{\tf\,\d\,b} -T^{\td\,\ \vf}_{\d\,b\,\tf}
H^{\tf\,\dg}_{\vf\,\tc\,a}
-T_{\tc\,b\,\df}^{\dg\,\ \tf} H^{\df\,\td}_{\tf\,\d\,a} \ =\ 0~.
\label{soldTH}
\eea
Although it contains a lot of information, this relation is less
complicated than it seems to be.
As it will be shown later on, the spinorial coefficients,
{\em i.e.} $H_{\g\,b\,a}^\tc$, $H_{\tc\,b\,a}^\dg$, $T_{\g\,b}^{\tc\,\ \au}$,
$T_{\tc\,b}^{\dg\,\ \au}$, $T_{\g\,\tb\,\da}^{\tc\,\db\,\ta}$,
$T_{\tc\,\b\,\ta}^{\dg\,\tb\,\a}$, $H_{\au \,b\,\a}^{\,\ \,\ \ta}$,
$H_{\au\,b\,\ta}^{\,\ \,\ \da}\,$ are all expressed in terms of one and the same
gravigino superfield (which contains the helicity 1/2 field of the multiplet in
its lowest component), as a consequence of the Bianchi identities at
dimension 1/2.
The soldering is achieved in requiring
\be
T_{\g\, \tb}^{\tc\, \db \, a} \ =\
-2i\,\d^\tc_\tb\,(\s^a\e)_\g{}^\db~,
\cem
H_{\g\, \tb \, a}^{\tc\, \db} \ =\
-2i\,L\,\d^\tc_\tb\,(\s_a\e)_\g{}^\db~,
\eqn{cons}
with the superfield $L$
pertaining to the real scalar of the theory. In the last
line of
(\ref{soldTH}) this gives rise to
\bea \lefteqn{ T_{\d\,\tc}^{\td\,\dg\,f} H_{f\,b\,a}
-T_{\d\,b\,\tf}^{\td\,\ \vf} H_{\vf\,\tc\,a}^{\tf\,\dg} -
T_{\tc\,b\,\df}^{\dg\,\
\tf} H_{\tf\,\g\,a}^{\df\,\td} \ =\  }\nonumber\\
&&-2i\,\d_\tc^\td\,(\s^f\e)_\d{}^\dg H_{fba}+2i\,L\,T_{\d\,b\,\tc}^{\td\,\
\g} \,
(\s_a\e)_\g{}^\dg +2i\,L\,T_{\tc\,b\,\dd}^{\dg\,\ \td} (\bs_a\e)^\dd{}_\d~.
\eea

This qualitative discussion
should give an idea in which way relations between $\,H_{c\,b\,a}$, $\cd_a
L$ and $T_{\g\,b\,\a}^{\tc\,\ \ta}$, $T_{\dg\,b\,\ta}^{\tc\,\ \da}\;$ will
arise.
It should also become clear that the full explicit analysis still requires a
certain amount of computational efforts. Observe that, as a consequence of the
above and other relations, $\,H_{c\,b\,a}\loco\,$ and $\,\cd_a L\loco\,$ will
appear in the supergravity transformation laws of
$\ \p_{m \, \ta}^{ \ \ \a}$,
$\ \bar{\p}_{m \, \da}^{\ \ \ta}$,
through $\,T_{\g\,b\,\ta}^{\tc\,\ \a}\loco\,$,
$\,T_{\dg\,b\,\da}^{\tc\,\ \ta}\loco$ , as identified in (\ref{dE_comp}).

In a similar way, relations between $\,T_{c\,b}{}^\au\,$, the graviphoton
fieldstrength, and $\,H_{\au\, b\,a}\,$ as well as $\,T_{\g\,b\,\da}^{\tc\,\
\ta}\,$, $\,T_{\tc\,b\,\ta}^{\dg\,\ \a}\,$ can be obtained. To illustrate
this we
consider the Bianchi identity $\brh{_{\d\g \, ba}^{\td\tc}}$ of
(\ref{BianchiH}) in
the 2--form sector:
\bea
&&\cd^\td_\d H^\tc_{\g\,b\,a}
+ \cd^\tc_\g H^\td_{\d\,b\,a}
\nn\\[2mm] &&{}
+T^{\td\,\tc\,\vf}_{\d\,\g\,\tf} H^\tf_{\vf\,b\,a}
+T^{\td\,\tc\,\tf}_{\d\,\g\,\df} H^\df_{\tf\,b\,a}
+T^{\td\,\ \au}_{\d\,a} H_{\au\,\g\,b}^{\,\ \tc}
+T^{\tc\,\ \au}_{\g\,a} H_{\au\,\d\,b}^{\,\ \td}
-T^{\td\,\ \au}_{\d\,b} H_{\au\,\g\,a}^{\,\ \tc}
-T^{\tc\,\ \au}_{\g\,b} H_{\au\,\d\,a}^{\,\ \td}
\nn\\[2mm] &&{}
+T^{\td\,\tc\,\au}_{\d\,\g} H_{\au\,b\,a}
+T_{b\,a}{}^\au H_{\au\,\d\,\g}^{\,\ \td\,\tc}
+T^{\td\,\ \tf}_{\d\,a\,\df} H^{\df\,\tc}_{\tf\,\g\,b}
+T^{\tc\,\ \tf}_{\g\,a\,\df} H^{\df\,\td}_{\tf\,\d\,b}
-T^{\td\,\ \tf}_{\d\,b\,\df} H^{\df\,\tc}_{\tf\,\g\,a}
-T^{\tc\,\ \tf}_{\g\,b\,\df} H^{\df\,\td}_{\tf\,\d\,a} \ =\ 0~.\nn
\eea
Again, we discard
for the moment the discussion of the derivative and quadratic spinor terms. The
terms pertinent for the soldering are those containing $T_{c\,b}{}^\au$,
$H_{\au\,b\,a}$ and $T_{\g\,b\,\da}^{\tc\,\ \ta}$. Here, as a second set of
soldering conditions, we take
\be
T_{\g\,\b}^{\tc\,\tb\,\au} \ =\  \e_{\g\b}\,
T^{[\tc\tb]\au}~,\cem H_{\au\,\b\,\a}^{\,\ \tb\,\ta} \ =\  \e_{\b\a}\,
H_\au{}^{[\tb\ta]}~,
\ee
and correspondingly for the complex conjugates. The relevant
terms in the above identity are
\be
T_{\g\,\b}^{\tc\,\tb\,\au}
H_{\au\, b\,a} +T_{b\,a}{}^\au H_{\au\,\d\,\g}^{\,\ \td\,\tc} \ =\
\e_{\d\g}\,\left(T^{[\td\tc]\au}H_{\au\, b\,a}+T_{b\,a}{}^\au
H_\au{}^{[\td\tc]}\right)~, \ee as well as \be T_{\d\,b\,\df}^{\td\,\ \tf}\,
H_{\tf\,\g\,a}^{\df\,\tc} \ =\ -2i\,L\,T_{\d\,b\,\dg}^{\td\,\ \tc}
\,(\bs_a\e)^\dg{}_\g~.
\ee
This should give an idea how, after some more
computational efforts, $T^{[\td\tc]\au} H_{\au ba}$ as well as
$T_{\g\,b\,\da}^{\tc\,\ \ta}$ will be related to $T_{c\,b}{}^\au$, the
graviphoton fieldstrength, with $T^{[\td\tc]\au} $ and $H_\au{}^{[\td\tc]}$
acting as converters between the central charge basis (indices $\au$, $\av$) and
the $SU(4)$ basis in the antisymmetric representation (indices $[\td\tc] $). A
more detailed analysis of the relevant Bianchi identities shows that
$T_{c\,b}{}^\au $ will appear in the tensor decomposition of the torsion
$T^{\td\,\ \tf}_{\d\,a\,\df}$ and in the derivative terms $\cd^\td_\d
H^\tc_{\g\,b\,a}$ as well. As will become clear in the explicit discussion,
$H^\tc_{\g\,b\,a}\loco$ will be related to the helicity 1/2 gravigini
components of
the multiplet. Therefore, the soldering operation implies that the
graviphoton field
strength appears in a well defined way in the supersymmetry transformation of
both the gravitini fields and the helicity 1/2 components.

On the other hand, the relation between $H_{\au\,b\,a}$ and $T_{c\,b}{}^\au$
implies the appearance of a graviphoton Chern--Simons form in $H_{c\,b\,a}
\loco$, as we will show now. Note that this is a property which appears at the
component field level. Using the double bar projection $E^\ca\doubar=e^\ca$
equivalently in the coordinate and the covariant frame basis, we obtain
\bea
H\doubar&=&\frac{1}{3!}dx^m dx^n dx^k \partial_k b_{nm}\\[3mm]
H\doubar&=&\frac{1}{3!}e^a e^b e^c H_{c\,b\,a}\loco
+\frac{1}{2}e^a e^b e^\au
H_{\au\,b\,a}\loco\nn\\[2mm]
&&+\frac{1}{2}e^a e^b e_\tc^\g
H^\tc_{\g\,b\,a}\loco +\frac{1}{2}e^a e^b e^\tc_\dg
H_{\tc\,b\,a}^\dg\loco\nn\\[2mm]
&&+\frac{1}{2} e^\au e_\tb^\b e_\tc^\g
H^{\tc\,\tb}_{\g\,\b\,\au}\loco +e^a e_\tb^\b e^\tc_\dg
H_{\tc\,\b\,a}^{\dg\,\tb}\loco +\frac{1}{2} e^\au e^\tb_\db e^\tc_\dg
H_{\tc\,\tb\,\au}^{\dg\,\db}\loco\nn \\[2mm]
&&+e^a e_\tb^\b e^\au
H_{\au\,\b\,a}^{\,\ \tb}\loco +e^a e^\tb_\db e^\au H_{\au\,\tb\,a}^{\,\
\db}\loco~.
\eea
The term giving rise to the Chern--Simons coupling of the
graviphotons is
\be
e^a e^b e^\au H_{\au\,b\,a}\loco \ = \ dx^mdx^ndx^le_l{}^a e_n{}^b
v_m{}^\au\,H_{\au ba}\loco~,
\eqn{CS}
due to the fact that
$e^\au=dx^m v_m{}^\au$ and that $H_{\au\, b\,a}$ is related to the graviphoton
field strength according to the previous discussion.

The qualitative discussion in this section should give a flavor of the
conceptual
basis of the central charge superspace and its impact on the structure of the
$N=4$ supergravity multiplet. In the sequel we provide some of the technical
details in a compact way.

\section{Superspace and the N--T multiplet}

As we have seen qualitatively, the soldering mechanism intertwining the 2--form
with the supergravity geometry is essential for the superspace description of
$N=4$ supergravity with an antisymmetric tensor. Let us now have a somewhat
closer
look at the superspace geometry pertaining to the N--T multiplet.

As has become common usage we shall exploit the consequences of torsion and
3--form curl constraints in analysing the respective Bianchi identities in the
order of increasing engineering dimension (in units of +1 for space--time and
+1/2 for spinor derivatives) normalized such that the torsion and 3--form curl
components displayed in (\ref{cons}) have dimension 0.

We shall however refrain here from a distinction between {\em constraints} and
{\em consequences of constraints} and simply display the properties of torsion
and 3--form curl suitable for the description of the N--T multiplet. We stress
first of all that the structure group is taken to be Lorentz$\times SU(4)\,$ in
the conventional superspace sector, and trivial in the central charge sector.

Vierbein and gravitini, as well as the spin connection being identified in the
usual way, we shall illustrate in the following the salient superspace structures
relevant for the description of the remaining components of the N--T multiplet.

\indent

\begin{itemize} \item {\em Graviscalar and gravigini} \end{itemize}

As to the
systematic discussion of $H_{\cc\cb\ca}\,$, we note that the coefficients at
dimension -1/2 (spinor indices only) all vanish. At dimension 0, see eq.
(\ref{cons}), we parametrize \be L \ =\  e^{2\phi}~. \ee The lowest component of
the real superfield $\phi$ will be identified as the scalar component field of
the N--T multiplet.

This real superfield will appear in the converters $T^{[\tc\tb]\au}$,
$T_{[\tc\tb]}{}^\au$ as well,
\be
T^{[\tc\tb]\au} \ =\
e^\phi\,a^{[\tc\tb]\au}~,\cem T_{[\tc\tb]}{}^\au \ =\
e^\phi\,a_{[\tc\tb]}{}^\au~.
\label{scal1}
\ee
Likewise, in the $H$--sector, we identify
\be
H_\au{}^{[\tc\tb]} \ =\  e^\phi\,m_\au{}^{[\tc\tb]}~,\cem H_{\au[\tc\tb]} \ =\
e^\phi\,m_{\au[\tc\tb]}~.
\label{scal2}
\ee
The constant matrices $a^{[\tc\tb]\au}$,
$a_{[\tc\tb]}{}^\au$ and $m_\au{}^{[\tc\tb]}$, $m_{\au[\tc\tb]}$ are
supposed to satisfy the
relations\footnote{We have fixed certain {\em a priori} free parameters such as
to fit with the N--T multiplet, other possibilities are studied in \cite{AMK}.}
\be
a^{[\td\tc]\au}\,m_\au{}^{[\tb\ta]} \ =\ 8\, \e^{\td\tc\tb\ta}~,
\cem
a_{[\td\tc]}{}^\au\, m_{\au[\tb\ta]}\ =\  8\,\e_{\td\tc\tb\ta}~,
\ee
\be
a^{[\td\tc]\au}\,m_{\au[\tb\ta]}+a_{[\tb\ta]}{}^\au\, m_\au{}^{[\td\tc]} \ =\
16\, \d^{\td\tc}_{\tb\ta}~.
\ee
Furthermore, we impose the
self--duality relations\footnote{In fact, these relations may be obtained as a
consequence of the particular structure of the superspace geometry
\cite{AMK}.}
\be
a^{[\td\tc]\au} \ =\  \frac12\, \e^{\td\tc\tb\ta}a_{[\tb\ta]}{}^\au~,\cem
m_\au{}^{[\tb\ta]} \ =\  \frac12\,m_{\au[\td\tc]}\e^{\td\tc\tb\ta}~. \ee
As a consequence, the second relation above becomes
\be
a^{[\td\tc]\au}\,m_{\au[\tb\ta]} \ =\  a_{[\tb\ta]}{}^\au\, m_\au{}^{[\td\tc]} \
=\ 8\,\d^{\td\tc}_{\tb\ta}~.
\ee
For consistency, we have as well
\be
m_{\au[\tf\te]}\, a^{[\tf\te]\av} \ =\ 16 \, \d^\av_\au~.
\ee
As a consequence, any
object $X_\au$ or $Y^\au$, once converted to the $SU(4)$ basis,
\bea &&
X^{[\td\tc]} \ =\  a^{[\td\tc]\au}\,X_\au~,\cem X_{[\tb\ta]} \ =\
a_{[\tb\ta]}{}^\au X_\au~,\\ && Y_{[\td\tc]} \ =\  Y^\au\, m_{\au[\td\tc]}~,\cem
Y^{[\tb\ta]} \ =\  Y^\au \,m_\au{}^{[\tb\ta]}~, \eea satisfies the self--duality
relations \be X^{[\td\tc]} \ =\  \frac12\, \e^{\td\tc\tb\ta} X_{[\tb\ta]}~,\cem
Y_{[\tb\ta]} \ =\  \frac12\,Y^{[\td\tc]}\e_{\td\tc\tb\ta}~,
\ee
similar to the
reality conditions employed in the description of the $N=4$ Yang--Mills theory
\cite{Soh78a}.

Whereas the real graviscalar is identified as the lowest component of the {\em
graviscalar superfield} $\phi$, its covariant fermionic partner,
the gravigino (in reference to the gaugino in
supersymmetric Yang--Mills theory), appears as
lowest component of the {\em gravigino superfield}
$\;T^{[\tc\tb\ta]}_{\ \ \ \ \da}$,
$T_{[\tc\tb\ta]}^{\ \ \ \a}$. Here, it is costumary to parametrize
\be
T^{[\tc\tb\ta]}_{\ \ \ \ \da} \ =\  \bl_{\da\td}\,\e^{\td\tc\tb\ta}~,\cem
T_{[\tc\tb\ta]}^{\ \ \ \a} \ =\  \la^{\a\td}\,\e_{\td\tc\tb\ta}~.
\ee
Recall that, at
dimension 1/2, there is quite a large number of torsion and 3--form curl
coefficients. Injecting the structure of torsion and 3--form at dimensions -1/2
and 0 into the dimension 1/2 Bianchi identities, together with certain
conventional
constraints, implies that all the non--vanishing coefficients are expressed
in terms
of the gravigino superfields. The coefficients given in terms of
$\la^{\a\ta}$ are
\bea
&& T_{\g\,\tb\,\da}^{\tc\,\db\,\ta} \ =\  -\frac14\,
\d^\db_\da\d_\tb^\ta\,\la_\g^\tc~, \cem T_{\g\,\b\,\a}^{\tc\,\tb\,\ta} \ =\
\frac14
(\d_\b^\a\d_\ta^\tb\,\la_\g^\tc+\d_\g^\a\d_\ta^\tc\,\la_\b^\tb)~,\nn\\[2mm]
&&\cem\mbox{and}\cem H_{\g\,b\,a}^\tc \ =\  -2(\s_{ba})_\g{}^\vf\,\la_\vf^\tc
\,e^{2\phi}~, \eea in the conventional superspace sector, whereas in the
central
charge sector we find
\be
T_{c\,\tb}^{\,\ \db\,\au} \ = \
\frac{i}{4}e^\phi\, \bs_c^{\db\vf}\,\la_\vf^\tf\,a_{[\tf\tb]}{}^\au~,
\cem
H_{\au\, b\,\ta}^{\,\ \,\ \da} \ = \
\frac{i}{4}e^\phi\, \bs_b^{\da\vf}\,\la_\vf^\tf\,m_{\au[\tf\ta]}~.
\ee
It is interesting to note that these two expressions are related through \be
H_{\au\,b\,\ta}^{\,\ \,\ \da} \ =\  T_{b\,\ta}^{\,\ \da\,\av}\,g_{\av\au} \cem
\mbox{with}\quad g_{\av\au} \ =\
\frac{1}{32}\,m_\av{}^{[\td\tc]}\,m_\au{}^{[\tb\ta]}\,\e_{\td\tc\tb\ta}~,
\ee
or, conversely
\be T_{b\,\ta}^{\,\ \da\,\au} \ = \
g^{\au\av}\,H_{\av\,b\,\ta}^{\,\ \,\ \da} \cem \mbox{with}\quad
g^{\au\av}\ = \
\frac{1}{32}\,a_{[\td\tc]}{}^\au a_{[\tb\ta]}{}^\av\eps^{\td\tc\tb\ta}~.
\ee
Of course, the metric $g_{\au\av}$ in the central charge basis satisfies
\be
g_{\au\aw}g^{\aw\av} \ = \ \delta_\au^\av~.
\ee
It should be clear that in deriving these results the
superspace soldering mechanism has already been at work.
Likewise, for the $\bl_{\da\ta}$ dependent coefficients, we find
\bea &&
T^{\dg\,\tb\,\a}_{\tc\,\b\,\ta} \ = \
-\frac14\,\d_\b^\a\d^\tb_\ta\, \bl^\dg_\tc~, \cem
T^{\dg\,\db\,\ta}_{\tc\,\tb\,\da} \ = \
\frac14 (\d^\db_\da\d^\ta_\tb\,\bl^\dg_\tc
+\d^\dg_\da\d^\ta_\tc\,\bl^\db_\tb)~,
\nn\\[2mm] && \cem \mbox{and} \cem
H^\dg_{\tc\,b\,a} \ = \
-2(\bs_{ba})^\dg{}_\df\,\bl^\df_\tc\,e^{2\phi}~, \eea
in the conventional sector, and
\be T_{c\,\b}^{\,\ \tb\,\au} \ = \
\frac{i}{4}e^\phi \, (\bs_c)_{\b\df}\,\bl^\df_\tf\,a^{\au[\tf\tb]}~, \cem
H_{\au\,b\,\a}^{\ \ \ \ta} \ = \
\frac{i}{4}e^\phi\, (\s_b)_{\a\df}\,\bl^\df_\tf\,m_\au{}^{[\tf\ta]}~,
\ee
in the central charge sector.
As before, $H_{\au\,b\,\a}^{\ \ \ \ta}$ and $T_{b\,\a}^{\,\ \ta\,\au}\,$ are
related through the metric in the central charge basis.

Finally, the dimension 1/2 Bianchi identities imply \be \la_\a^\ta \ =\  -2\,
\cd_\a^\ta\phi~,\cem \bl^\da_\ta \ =\
 -2\, \bar\cd_\ta^\da\phi~. \ee We have thus
completely exhausted the information contained in the dimension 1/2 Bianchi
identities.

\indent

\begin{itemize} \item {\em Graviphotons} \end{itemize}

Central charge superspace has been conceived as a concise framework to describe
graviphotons as messengers of local central charge transformations based on a
sound geometric basis. The covariant fieldstrength of the graviphotons is
identified in the superspace torsion coefficient $T_{c\,b}{}^\au\,$.

As anticipated in the introduction, the superspace soldering procedure
implies that
certain torsion and 3--form curl coefficients will be expressed in terms of the
graviphoton torsion, intertwining supergravity and the 2--form geometries in an
intricate way. As to the torsion coefficients, at dimension 1 we find \bea
T_{\g\, b\, \da}^{\tc\ \, \ta} &=&\frac{i}{8}\,e^{-\phi}\,(\s_b\bs^{dc})_{\g\,
\da}\,F_{dc}{}^{[\tc\ta]}~,\\ T_{\tc\, b\, \ta}^{\dg\ \, \a} &=&
\frac{i}{8}\,e^{-\phi}\,(\bs_b\s^{dc})^{\dg\,\a}\,F_{dc\,[\tc\ta]}~, \eea where
we have introduced the notation \be F_{dc}{}^{[\tb\ta]} \ = \ T_{dc}{}^\au
m_\au{}^{[\tb\ta]}~,\cem F_{dc\,[\tb\ta]} \ = \ T_{dc}{}^\au m_{\au\,[\tb\ta]}~.
\ee These relations determine the appearance of the graviphotons in the
supersymmetry transformations laws of the gravitini.

As a consequence of the superspace soldering, the graviphotons appear in the
2--form geometry in the coefficients
\be
H_{\au\, b\, a} \ = \ T_{b\,a}{}^\av\,g_{\av\au}~.
\ee
As we have already stressed, this is responsible
for the appearance of graviphoton Chern--Simons forms in the supercovariant curl
of the antisymmetric tensor. Finally, the graviphoton fieldstrength appears
also in the second spinor derivative of the graviscalar superfield such that
\be
\cd_\b^\tb\cd_\a^\ta \phi \ = \
- \frac{1}{4} \, e^{-\phi} (\s^{ba}\e)_{\b\a}\,F_{ba}{}^{[\tb\ta]}
- \frac{3}{8} \, \la_\b^\tb \la_\a^\ta~,
\ee
\be
\cd_\tb^\db \cd_\ta^\da \phi \ = \
- \frac{1}{4} \, e^{-\phi} (\bs^{ba}\e)^{\db\da}\,F_{ba}{}_{[\tb\ta]}
- \frac{3}{8} \, \bl_\tb^\db \bl_\ta^\da~.
\ee
This last relation
illustrates how $F_{ba}{}^{[\tb\ta]}$ appears in the supersymmetry
transformations of
the gravigini.

\indent

\begin{itemize} \item {\em Antisymmetric tensor} \end{itemize}

As to the antisymmetric tensor $\,b_{mn}$, the superspace soldering
mechanism allows to identify its covariant 3--form fieldstrength
$H_{cba}$ in certain torsion coefficients. More explicitly,
as a consequence of our constraints, the Bianchi identities
give rise to
\bea
T^{\tc \ \, \a}_{\g \, b \, \ta} &=&
-\frac{1}{4} \, \d ^\tc_\ta \, \d^\a_\g \, H^*_b e^{-2\f}
-\frac{i}{8} \left( \d^\tc_\ta \, \la^\tf_\g \bl_{\da \tf} \, \bs_b^{\da \a}
+ 2 \, \la^{\a \tc} \s_{b \, \g \dg} \bl_\ta^\dg \right)~,
\label{at1}\\
T_{\tc \, b \, \da}^{\dg \ \, \ta} &=&
-\frac{1}{4} \, \d _\tc^\ta \, \d_\da^\dg \, H^*_be^{-2\f}
-\frac{i}{8}\left(\d_\tc^\ta \, \bl_\tf^\dg \la^{\a \tf} \s_{b \, \a \da}
+ 2  \, \bl_{\da \tc} \bs_b^{\dg \g} \la^\ta_\g \right)~,\label{at2}
\eea
with the dual tensor defined as
\be
H^*_d=\frac{1}{3!}\,\e_{dcba}\,H^{cba}~.
\ee
This indicates how the antisymmetric tensor will appear in the supersymmetry
transformation of the gravitino, after suitable identification and projection to
lowest components in (\ref{dE_comp}).

In turn, the same Bianchi identity leads to the relation
\be
\cd^\db_\tb \cd_\a^\ta \phi \ = \
i \d_\tb^\ta (\s^a \e)_\a {}^\db \cd_a \phi
+ \frac{1}{2} e^{-2\phi} \d_\tb^\ta (\s^a \e)_\a {}^\db H_a^*
-\frac{3}{8}
\left( \la_\a^\ta \bl_\tb^\db -2 \d_\tb^\ta \la_\a^\tf \bl_\tf^\db \right)~,
\ee
illustrating how the antisymmetric tensor appears in the supersymmetry
transformation of the gravigini.

\section{Component field transformations}

As to  the component fields of the N--T multiplet, the graviton,
gravitini and graviphotons are identified in the superspace
frame $E^\ca$ \equ{frame}, the antisymmetric tensor in the superspace
2--form $B$ \equ{B},
whereas the real scalar and the helicity 1/2 gravigini fields are found as
the lowest  components of the superfields
$\f$, $\la_\a^{\ta}$, $\bl_\ta^\da$.
We shall use the same symbols to denote the component fields, {\em i.e.}
\be
\f\loco \ = \ \f~, \hspace{1cm}
\la_\a^{\ta}\loco \ = \ \la_\a^{\ta}~, \hspace{1cm}
\bl_\ta^\da\loco \ = \ \bl_\ta^\da~.
\ee
The Wess-Zumino transformations of graviton,
gravitini and graviphotons can then be read off from suitable projections
of the superfield equation (\ref{dE_comp}). For the frame in space-time one
obtains the usual result
\be
\d_\wz \, e_{m}{}^{a} \ = \
-i \, \p_{m \, \tb}^{\ \ \b}
(\sigma^a \e)_\b{}^{ \db} \, \bar{\zeta}^{\tb}_\db
-i \, \bar{\psi }_{m \, \db}^{\ \ \tb}
        (\bar{\sigma }^a \e)^\db{}_\b \, \zeta_\tb^\b~,
\ee
whereas the transformations for the Rarita-Schwinger fields are
\bea
\d_\wz \, \p_{m \, \ta}^{\ \ \a} &=&
2 \, \cd_{\! m} \, \zeta_\ta^\a
-\frac{i}{4}e^{-\phi}e_m{}^b (\bar{\zeta}^\tb\bar{\sigma}^{a})^{\alpha }
 F^+_{ba \, [\tb\ta]} \loco
-\frac{1}{2} e^{-2\phi }e_{m}{}^{b} \zeta_\ta^\a H_{b}^{*}\loco
\nn \\ &&
+ \, \frac{1}{4} \, \p_{m \, \ta}^{\ \ \a}
(\zeta \cdot \la - \bar{\zeta} \cdot \bl)
+ \frac{1}{4} (\psi_m \cdot \la - \bar \psi_m \cdot \bl ) \, \zeta_\ta^\a
\nn \\ &&
+ \, \frac{i}{2} \la^{\a \, \tb}(\zeta_\tb\sigma_m\bl_\ta
   -2i \bar{\psi}_m{}^\td \bar{\zeta}^{\tc} \varepsilon _{\td\tc\tb\ta})
-\frac{i}{4}(\zeta_\ta \la^\tb)( \bl_\tb \bar \sigma_m)^\a~,
\eea
\bea
\d_\wz \, \bar{\psi}_{m \, \da}^{\ \ \ta} &=&
2 \, \cd_{\! m} \, \bar{\zeta}_\da^\ta
-\frac{i}{4}e^{-\phi }e_m{}^b (\zeta_\tb \sigma^a)_\da
F^-_{ba}{}^{[\tb \ta]}\loco
+\frac{1}{2}e^{-2\phi}e_m{}^b \bar{\zeta}_\da^\ta H_b^* \loco
\nn \\ &&
- \, \frac{1}{4} \, \bar{\psi}_{m \, \da}^{\ \ \ta}
(\zeta \cdot \la - \bar{\zeta} \cdot \bl)
- \frac{1}{4} (\psi_m \cdot \la - \bar \psi_m \cdot \bl ) \, \bar \zeta^\ta_\da
\nn \\ &&
+ \, \frac{i}{2} \bl_{\da \tb}
(\bar{\zeta}^\tb\bar{\sigma}_{m}\la^\ta
- 2i \psi_{m}{}_\td \zeta_\tc \varepsilon^{\td\tc\tb\ta})
+\frac{i}{4}(\bar{\zeta}^\ta \bl_\tb) (\la^\tb \sigma_m)_\da~,
\eea
with definitions
$F^\pm_{ba} = F_{ba} \pm \frac{i}{2}F^{dc} \, \varepsilon_{dcba}$
for the selfdual and antiselfdual components of the central charge fieldstrength
and
$\zeta \cdot \la = \zeta^\a_\ta \la_\a^\ta$,
$ \ \bar{\zeta} \cdot \bl = \bar{\zeta}_\da^\ta \bl^\da_\ta$
for summation conventions.
Observe also the appearance of the field strength of the antisymmetric tensor,
which is due to the soldering mechanism. As the supervielbein contains the
graviphotons as well, relation (\ref{dE_comp}) implies
\be
\d_\wz V_{m}{}^{[{\tb}{\ta}]} \ = \
- 2 e^\phi (i \zeta_\td \sigma_m\bl_\tc
- 2 \zeta_\td \psi_{m\, \tc}) \, \e^{\td\tc\tb\ta}
-2 e^\phi(i\bar{\zeta}^\td \bar{\sigma}_m \la^\tc
-2 \bar{\zeta}^\td \bar{\psi}_m{}^\tc) \, \d^{\tb\ta}_{\td\tc}~.
\ee
As to the antisymmetric tensor one applies (\ref{dB_comp}), with the result
\bea
\d_\wz \, b_{mn} &=&
-2 \, e^{2 \phi}(\zeta_\ta \sigma_{mn} \la^\ta)
-2 \, e^{2 \phi}(\bar{\zeta}^\ta \bar{\sigma}_{mn}\bl_\ta)
-i(\bar{\psi}_{[m}{}^\ta\bar{\sigma}_{n]}\zeta_\ta)
-i(\psi_{[ m\ta}\sigma_{n ]}\bar{\zeta}^\ta)
\nn \\ &&
+\frac{1}{2} V_{[m}{}^{[{\tb}{\ta}]}(\psi_{n ] \ta}\zeta_{\tb})
-\frac{i}{4} e^\phi \, V_{[m}{}^{[{\tb}{\ta}]}
      (\zeta_\ta \sigma_{n ]}\bl_\tb)
\nn \\ &&
+\frac{1}{2} V_{[m \, [{\tb}{\ta}]} (\bar{\psi}_{n ]}{}^\ta \bar{\zeta}^\tb)
-\frac{i}{4} e^\phi \, V_{[m \, [{\tb}{\ta}]}
     (\bar{\zeta}^\ta \bar{\sigma}_{n]}\la^\tb)~,
\eea
where the antisymmetrization convention $X_{[m} Y_{n]} = X_m Y_n - X_n Y_m$
for any
vectors $X_m$ and $Y_n$ is used. Finally, the transformations of graviscalar and
gravigini are derived in the usual way as well, one obtains
\be
\d_\wz \, \phi \ = \ \zeta_\ta^\a \la_\a^\ta
+\bar{\zeta} _\da^\ta \bl_\ta^\da~,
\ee
\bea
\d_\wz \, \la_\a^{\ta} &=&
-2i (\bar{\zeta}^\ta \bar{\sigma}^a \e )_\a \, \cd_{\! a} \phi \loco
+\frac{1}{2}e^{-\phi }(\zeta_\tb\sigma^{ba}\e)_\a
\, F_{ba}{}^{[{\tb}{\ta}]} \loco
-e^{-2\phi }(\bar{\zeta}^\ta \bar{\sigma}^a \e)_\a
H_{a}^{*} \loco
\nn \\ &&
+ 3  (\zeta \cdot \la - \bar{\zeta} \cdot \bl) \la_\a^\ta
+6 (\bar{\zeta}^\ta \bl_\tb) \la_\a^\tb~,
\eea
\bea
\d_\wz \, \bl_\ta^\da &=&
-2i (\bar{\zeta}^\ta \bar{\sigma}^a \e )_\a \, \cd_{\! a} \phi \loco
+\frac{1}{2} e^{-\phi }(\bar{\zeta}^\tb \bar{\sigma}^{ba} \e)^\da
F_{ba}{\, }_{[\tb\ta]}\loco
+e^{-2\phi }(\zeta_\ta \sigma^a \e )^\da H_a^* \loco
\nn \\ &&
-3  (\zeta \cdot \la - \bar{\zeta} \cdot \bl) \bl_\ta^\da
+6 (\zeta_\ta \la^\tb) \bl_\tb^\da~.
\eea
In these expressions the component field supercovariant field strength and
derivatives are defined as
\bea
F_{ba}{}^{[\tb\ta]} \loco &=&
e_b{}^n e_a{}^m \partial_{[ n} V_{m ]} {}^{[\tb\ta]}
-2 e_b{}^n e_a{}^m \, e^\phi
\left(
 \p_{m \, \td} \p_{n\, \tc}\e^{\td\tc\tb\ta}
+ \bp_m{}^\td\bp_n{}^\tc \d^{\tb\ta}_{\td\tc}
\right)
\nn \\ &&
+i \, e_b{}^n e_a{}^m \, e^\f \left(
 \p_{[ m \, \td} \s_{n ]} \bl_\tc \e^{\td\tc\tb\ta}
+\bp_{[ m}{}^\td \bs_{n ]} \la^\tc \d^{\tb\ta}_{\td\tc}
\right)~,
\eea
for the graviphotons,
\bea
H^{*b}\loco &=& \frac{1}{2} \left(
\partial_m b_{lk}
-\frac{1}{8}V_{m [\tb\ta]} \partial_l V_k{}^{[\tb\ta]}
\right)
\e^{klmn} e_n{}^b
\nn \\ &&
 \frac{1}{2} e^{2 \f} \left(
i \p_{m \, \tb}\s_l \bp_k{}^\tb
+ \p_{m\, \tc}\s_{lk} \la^\tc
+ \bp_m{}^\tc \bs_{lk} \bl_\tc
\right)
\e^{klmn} e_n{}^b~,
\eea
for the antisymmetric tensor and
\be
\cd_a \phi \loco \ = \ e_a{}^m \partial_m \phi
+ \frac{1}{4} e_a{}^m (\psi_m \cdot \la - \bar \psi_m \cdot \bl)~,
\ee
for the graviscalar.

As already stressed, central charge transformations are special cases of
supergravity transformations as well. The central charge transformations of
component fields are defined in terms of local parameters
$\x^\ca\loco = (0,0,0,\z^\au)$. Assuming that the dependence of the
component fields
on the central charge directions is trivial, it is immediate to see that
only the
graviphotons and the antisymmetric tensor have non--trivial central charge
transformations.
Using again \equ{dE_comp} and \equ{dB_comp} with suitable component field
projections, one derives the transformation laws
\be
\d_{\bf c c} \, V_m{}^{[\tb\ta]} \ = \ \partial_m \zeta^{[\tb\ta]}~,
\ee
for the graviphotons and
\be
\d_{\bf c c} \, b_{mn} \ = \ \frac{1}{16} \left(
\partial_m V_n{}^{[\tb\ta]} - \partial_n V_m{}^{[\tb\ta]}
\right) \zeta_{[\tb\ta]}~,
\ee
for the antisymmetric tensor with the obvious notation
$\zeta^{[\tb\ta]} = \zeta^\au m_\au{}^{[\tb\ta]}$.

\section{Conclusion}

We have shown that the identification of the graviphotons in
$E_\cm{}^\ca$, the frame of superspace \cite{How82},
together with the soldering mechanism, which provides an intimate relation
between  gravity and 2--form geometries, clarifies a number of points in the
description of the N--T supergravity, {\em i.e.} $N=4$ supergravity with an
antisymmetric tensor gauge field.

One of the important points is that the graviscalar appears both in the
gravity and
in the 2--form sectors, as given in (\ref{scal1}) and (\ref{scal2}).
The symbols $a^{\au[\td\tc]}$ and $m_\au{}^{[\td\tc]}$ which appear there
serve to convert between the central charge and the $SU(4)$ bases. Moreover, the
particular constraint structure in superspace suggests selfduality relations
\be
a^{\au[\td\tc]} \ = \
\frac{1}{2} \e^{\td\tc\tb\ta} a^\au{}_{[\tb\ta]}~,
\qquad \qquad
m_\au{}^{[\td\tc]} \ = \
\frac{1}{2}\e^{\td\tc\tb\ta} m_{\au[\tb\ta]}~.
\ee
As a consequence, quantities as the graviphoton field strength
\be
F_\cd{}_\cc{}^{[\tb\ta]} \ = \
T_\cd{}_\cc{}^\au m_\au{}^{[\tb\ta]} \ = \
a^{[\tb\ta]}{}^\au H_\au{}_\cd{}_\cc~,
\ee
are selfdual in the same sense. The appearance of the graviphoton field
strength in
the 2--form sector explains in a neat way the presence of the graviphoton
Chern--Simons form in the supercovariant curl $H_{cba} \loco$of the
antisymmetric
tensor. It should be interesting to reinvestigate the coupling of $N=4$
matter to
$N=4$ supergravity \cite{Cha81}, \cite{dR85} in the geometric framework set
up here
as well.


\begin{thebibliography}{10}

\bibitem{Das77}
A.~Das.
\newblock {SO}(4) invariant extended supergravity.
\newblock {\em Phys. Rev.}, D15:2805--2809, 1977.

\bibitem{CS77}
E.~Cremmer and J.~Scherk.
\newblock Algebraic simplifications in supergravity theories.
\newblock {\em Nucl. Phys.}, B127:259--268, 1977.

\bibitem{CSF77}
E.~Cremmer, J.~Scherk, and S.~Ferrara.
\newblock {U(N)} invariance in supergravity theories.
\newblock {\em Phys. Lett.}, 68B:234--268, 1977.

\bibitem{CSF78}
E.~Cremmer, J.~Scherk, and S.~Ferrara.
\newblock {SU}(4) invariant supergravity theory.
\newblock {\em Phys. Lett.}, 74B:61--64, 1978.

\bibitem{FS78}
D.~Z. Freedman and J.~H. Schwarz.
\newblock N=4 supergravity theory with local {SU(2)xSU(2)} invariance.
\newblock {\em Nucl. Phys.}, B137:333--337, 1978.

\bibitem{Gri78}
M.~T. Grisaru.
\newblock Anomalies, field transformations, and the relation between {SU(4)}
  and {SO(4)} supergravity.
\newblock {\em Phys. Lett.}, 79B:225--230, 1978.

\bibitem{GSO77}
F.~Gliozzi, J.~Scherk, and D.~Olive.
\newblock Supersymmetry, supergravity theories and the dual spinor model.
\newblock {\em Nucl. Phys.}, B122:253--290, 1977.

\bibitem{NT81}
H.~Nicolai and P.K. Townsend.
\newblock {N}=3 supersymmetry multiplets with vanishing trace anomaly: building
  blocks of the {N}$>$3 supergravities.
\newblock {\em Phys. Lett.}, 98B:257--260, 1981.

\bibitem{Sie81}
W.~Siegel.
\newblock On-shell {0(N)} supergravity in superspace.
\newblock {\em Nucl. Phys.}, B177:325--332, 1981.

\bibitem{How82}
P.~Howe.
\newblock Supergravity in superspace.
\newblock {\em Nucl. Phys.}, B199:309--364, 1982.

\bibitem{Gat83}
S.J. Gates.
\newblock On-shell and conformal {N}=4 supergravity in superspace.
\newblock {\em Nucl.Phys.}, B213:409--444, 1983.

\bibitem{GD89}
S.J. Gates and J.W. Durachta.
\newblock Gauge two-form in {D}=4, {N}=4 supergeometry with {SU}(4)
  supersymmetry.
\newblock {\em Mod. Phys. Lett.}, A4:2007--2016, 1989.

\bibitem{WB83}
J.~Wess and J.~Bagger.
\newblock {\em Supersymmetry and Supergravity}.
\newblock Princeton Series in Physics. Princeton University Press, Princeton,
  1983.
\newblock 2nd edition 1992.

\bibitem{BGG00}
P.~Bin\'etruy, G.~Girardi, and R.~Grimm.
\newblock {\em Supergravity couplings: a geometric formulation}.
\newblock hep-th/0005225, to appear in Phys. Rep.

\bibitem{AMK}
A.~Kiss.
\newblock Ph.D. Thesis, to appear.

\bibitem{Soh78a}
M.~Sohnius.
\newblock Bianchi identities for supersymmetric gauge theories.
\newblock {\em Nucl.Phys.}, B136:461--474, 1978.

\bibitem{Cha81}
A.~H. Chamseddine.
\newblock N=4 matter coupled to {N}=4 matter and hidden symmetries.
\newblock {\em Nucl. Phys.}, B185:403--415, 1981.

\bibitem{dR85}
M.~de~Roo.
\newblock Matter coupling in {N}=4 supergravity.
\newblock {\em Nucl. Phys.}, B255:515--531, 1985.

\end{thebibliography}
\end{document}